   \documentclass[structabstract]{aa}

\usepackage{graphicx}
\usepackage[]{natbib}
\usepackage{txfonts}
\usepackage{booktabs}
\usepackage{url}
\usepackage{xcolor}
\usepackage{sidecap}
\usepackage{amssymb}
\bibpunct{(}{)}{;}{a}{}{,} 
\usepackage[normalem]{ulem}

\begin{document}

\title{The bimodal [Mg/Fe] versus [Fe/H] bulge sequence as revealed by APOGEE DR14}

    
\author{A.~Rojas-Arriagada \inst{\ref{inst1},\ref{inst2}}
  \and M.~Zoccali \inst{\ref{inst1},\ref{inst2}} 
  \and M.~Schultheis \inst{\ref{inst3}}
  \and A.~Recio-Blanco \inst{\ref{inst3}}
  \and G.~Zasowski \inst{\ref{inst4}}
  \and D.~Minniti \inst{\ref{inst2},\ref{inst5},\ref{inst6}}
  \and H.~J\"{o}nsson \inst{\ref{inst7}}
  \and R.~E.~Cohen  \inst{\ref{inst8},\ref{inst9}}
}

\institute{
Instituto de Astrof\'{i}sica, Facultad de F\'{i}sica, Pontificia Universidad Cat\'olica de Chile, Av. Vicu\~na Mackenna 4860, Santiago, Chile  \email{arojas@astro.puc.cl} \label{inst1}
\and
Millennium Institute of Astrophysics, Av. Vicu\~{n}a Mackenna 4860, 782-0436 Macul, Santiago, Chile \label{inst2}
\and
Laboratoire Lagrange, Universit\'e C\^ote d'Azur, Observatoire de la C\^ote d'Azur, CNRS, Bd de l'Observatoire, CS 34229, 06304 Nice cedex 4, France \label{inst3}
\and
Department of Physics \& Astronomy, University of Utah, Salt Lake City, UT, 84112, USA \label{inst4}
\and
Departamento de Ciencias F\'isicas, Universidad Andr\'es Bello, Campus La Casona, Fern\'andez Concha 700, 1058 Santiago, Chile \label{inst5}
\and
Vatican Observatory, 00120 Vatican City State, Italy \label{inst6}
\and
Lund Observatory, Department of Astronomy and Theoretical Physics, Lund University, Box 43, SE-22100 Lund, Sweden \label{inst7}
\and
Departamento de Astronom\'ia, Universidad de Concepci\'on, Casilla 160-C, Concepci\'on, Chile \label{inst8}
\and
Space Telescope Science Institute, 3700 San Martin Drive, Baltimore, MD 21218, USA \label{inst9}
}

   \date{Received...; accepted...}

   \newcommand{\teff}{${\rm T_{eff}}$}
   \newcommand{\logg}{$\log{g}$}
   \newcommand{\feh}{$\rm [Fe/H],$}
   \newcommand{\met}{${\rm [M/H]}$}
   \newcommand{\alfafe}{${\rm [\alpha/Fe]}$}
   \newcommand{\kms}{km~s$^{-1}$}
   \newcommand{\vrad}{${\rm V_{rad}}$}

 
  \abstract
   {The Galactic bulge has a bimodal metallicity distribution function: different kinematic, spatial, and, potentially, age distributions characterize the metal-poor and metal-rich components. Despite this observed dichotomy, which argues for different formation channels for those stars, the distribution of bulge stars in the $\alpha$-abundance versus metallicity plane has been found so far to be a rather smooth single sequence.}
   {We use data from the fourteenth data release of the APOGEE spectroscopic survey (DR14) to investigate the distribution in the Mg abundance (as tracer of the $\alpha$-elements)-versus-metallicity plane of a sample of stars selected to be in the inner region of the bulge.}
   {A clean sample has been selected from the DR14 using a set of data- and pipeline-flags to ensure the quality of their fundamental parameters and elemental abundances. An additional selection made use of computed spectro-photometric distances to select a sample of  likely bulge stars as those with ${\rm R_{GC}\leq 3.5 kpc}$. We adopt magnesium abundance as an $\alpha$-abundance proxy for our clean sample as it has been proven to be the most accurate $\alpha$-element as determined by ASPCAP, the pipeline for data products from APOGEE spectra.}
   {From the distribution of our bulge sample in the [Mg/Fe]-versus-[Fe/H] plane, we found that the sequence is bimodal. This bimodality is given by the presence of a low-Mg sequence of stars parallel to the main high-Mg sequence over a range of $\sim 0.5$~dex around solar metallicity. The two sequences merge above ${\rm [Fe/H]\sim0.15~ dex}$ into a single sequence whose dispersion in [Mg/Fe] is larger than either of the two sequences visible at lower metallicity. This result is confirmed when we consider stars in our sample that are inside the bulge region according to trustworthy Gaia DR2 distances.}
   {}

   \maketitle
   
\section{Introduction}
\label{sec:introduction}
Our understanding of the rather complex nature of the Galactic bulge has substantially increased in the last decade. Recent large-scale photometric and high-resolution spectroscopic surveys have turned their eyes towards the bulge, producing samples of optical and near-infrared (NIR) data of   unprecedented size. These large datasets have been complemented by several smaller projects obtaining pencil beam samples in low-extinction regions. In general terms, all the evidence provided by the different surveys and individual works argues for a dual nature of the bulge. Already seen from the bimodality (if not a multimodality) in the metallicity distribution function (MDF), the bulge seems to be the product of the spatial coexistence of (at least) two groups of metal-rich and metal-poor stars, with different kinematics \citep{babusiaux2010,ness2013b,zasowski2016}, spatial distributions \citep{zoccali2017}, $\alpha$-enhancements \citep{gonzalez2015,rojas-arriagada2017}, and seemingly different ages \citep{bensby2017,schultheis2017}.
All things considered, there is a consensus in the community concerning the origin of the metal-rich bulge. These stars have their origin in the secular evolution of the early thin disk, through bar formation, buckling and the consequent redistribution of angular momentum in the vertical direction. This is consistent with the X-shaped spatial distribution of these stars, as revealed by the bimodal red clump (RC)  spatial variations \citep{mcwilliam2010,saito2011,wegg2013}, their line-of-sight radial velocity distributions revealing the predominance of resonant orbit families \citep{williams2016}, and their younger ages \citep{schultheis2017}.

On the other hand, there is ongoing debate over the origin of the metal-poor bulge \citep{barbuy2018}. It has been proposed to be the product of an in-situ early violent process of star formation \citep{hill2011,rojas-arriagada2014}, or the product of the secular evolution of the early thick disk \citep{dimatteo2015,debattista2017}.

In spite of all the evidence suggesting a different nature for the metal-rich and metal-poor bulge stars, no clear discontinuity in the  $\alpha$-abundance-versus-metallicity plane has been found so far, with only some indication for a small vertical shift in the sequence around solar metallicity \citep{hill2011}. Such a break in the chemical sequence would be an important signature pinpointing the presence of two populations with different origins coexisting in the central kiloparsecs of the Milky Way.

In this work, we present observational evidence for the existence of a bimodal bulge sequence with a gap in $\alpha$-abundance at around solar metallicity. In this way we use chemistry to reinforce the evidence for the existence of two formation channels for stars currently located in the bulge region.

In Sect.~\ref{sec:data} we present a general description of the selected APOGEE data. The computed distances, the specific selections made to define our working samples, and an overview of the vertical variation of the bulge MDF are presented in Sect.~\ref{sec:sample_selections}. The double bulge sequence in the $\alpha$-abundance-versus-metallicity plane is presented and characterized in Sect.~\ref{sec:double_seq}. Finally, the discussion and our conclusions are presented in Sect.~\ref{sec:discussions_conclusions}.

\section{The data}
\label{sec:data}
The Apache Point Observatory Galactic Evolution Experiment \citep[APOGEE, ][]{majewski2017} survey started as one of the four Sloan Digital Sky Survey (SDSS) III experiments \citep{eisenstein2011,blanton2017} with the main aim being to perform a massive chemical cartography of the Milky Way stellar populations. Observations are performed at bright time with a dedicated 300-fiber, cryogenic, high-resolution ($R\sim22500$) spectrograph working in the NIR H band \citep[1.51-1.70 $\mu$m][]{wilson2019} mounted on the SDSS 2.5 m telescope \citep{gunn2006}. By observing at NIR wavelengths, APOGEE overcomes the limitations imposed by the large amount of dust extinction in the Galactic plane and towards the bulge, which have historically limited optical studies in these inner areas. The main targets of APOGEE are giant stars (RGB, AGB and RC), which are intrinsically luminous tracers present in stellar populations of all ages. They are selected by adopting 2MASS photometry and are observed from $H\sim7$ down to a flux limit of nearly $H=13.8$. An increasingly large sample of high-resolution and high signal-to-noise ratio (S/N, typically larger than 100 per resolution element) has been obtained for a large number of stars, up to $\sim277~000$ in the current fourteenth data release \citep[hereafter DR14,][]{holtzman2018}. The consortium has developed an observing strategy \citep{zasowski2013,zasowski2017}, a pipeline for data reduction and calibration \citep{nidever2015}, and a pipeline (the ASPCAP pipeline) which produces, in an automatic and homogeneous way, fundamental stellar parameters and abundances of 24 elements ($\alpha$-elements, iron-peak and Z-odd) at up to 0.1~dex precision \citep{garcia-perez2016,holtzman2018}. 

\begin{figure}[]
\centering
\includegraphics[width=9.2cm]{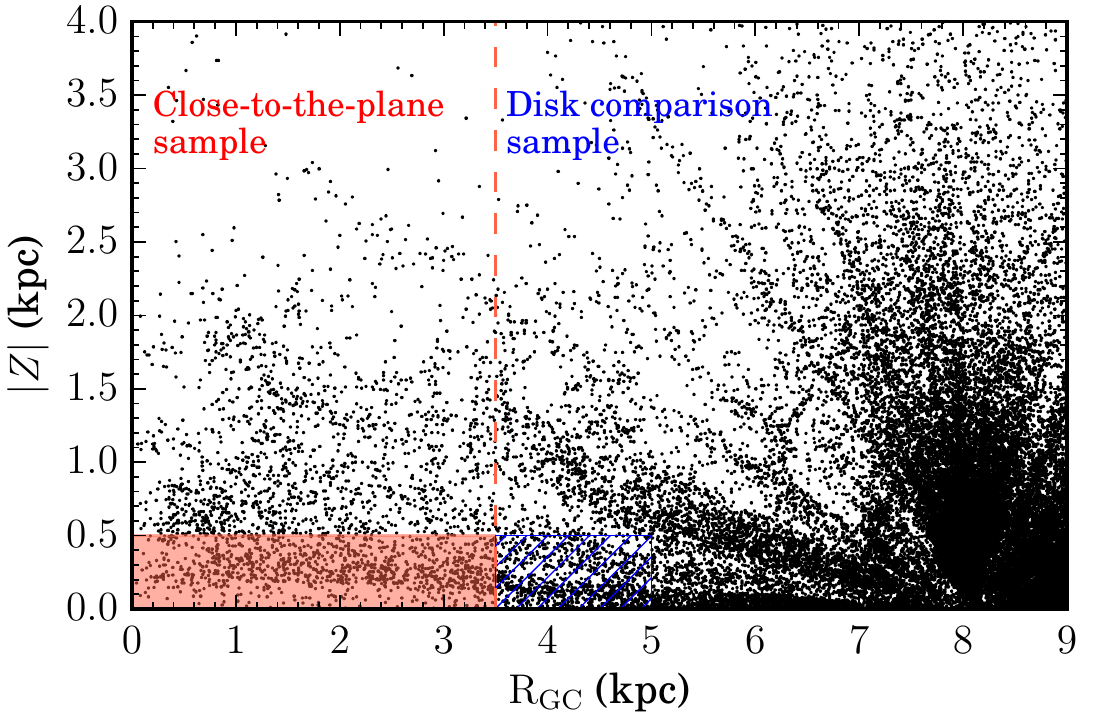}
\includegraphics[width=6.3cm]{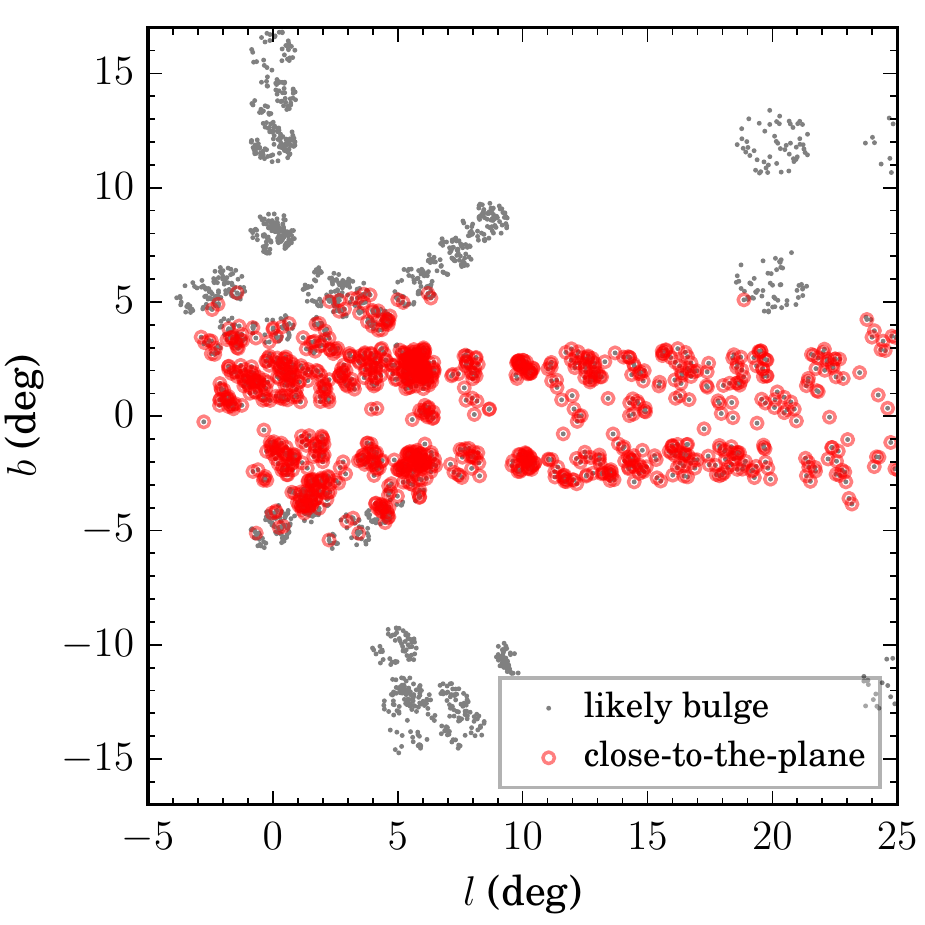}
\caption{Spectro-photometric distances. \textit{Upper panel:} Distribution of stars in radial Galactocentric distance vs. absolute distance from the plane. Black points depict the flag-culled sample selected from DR14 as described in the main text. The dashed vertical line and the shaded area depict the cuts used to select our likely bulge sample and the close-to-the-plane subselection, respectively. The hatched area highlights the region used to select a disk-comparison sample (Sect.~\ref{subsec:kinematics}). \textit{Lower panel:} $(l,b)$ distribution of the likely bulge (gray dots) and close-to-the-plane (red circles, with transparency to emphasize their concentration towards the center) samples.}
\label{fig:distances}
\end{figure}

In this work, we made use of data released as part of the APOGEE-DR14 \citep[SDSS Data Release 14,][]{abolfathi2018}. We adopt the set of fundamental parameters, effective temperature (\teff), and surface gravity (\logg), which are astrophysically calibrated relative to independent measurements in a calibration dataset (through the IRFM method and asteroseismology, respectively), and global metallicity (\met) which is calibrated using a set of globular clusters.

We made use of the set of APOGEE spectroscopic flags\footnote{\url{http://www.sdss.org/dr14/algorithms/bitmasks/}} to select a sample of stars with reliable parameter determinations. These different flags provide a complete description of the targeting status of the observed object (main survey sample, commissioning, calibration, and ancillary programs), cosmetics and/or reduction issues with the raw data, and the quality of the fitting procedure producing the set of fundamental parameters and elemental abundances. We refer to \citet{holtzman2015} for a detailed description of the APOGEE flags. Our selection is similar, but is more stringent than  those adopted in other APOGEE works studying field stellar populations \citep[e.g.,][]{jonsson2018,zasowski2019}. In particular, we use TARGFLAGS to select main survey field stars, STARFLAGS\footnote{Excluding stars with any of the flags: BAD\_PIXELS, BAD\_RV\_COMBINATION, BRIGHT\_NEIGHBOR, COMMISSIONING, PERSIST\_HIGH, PERSIST\_MED, PERSIST\_JUMP\_NEG, PERSIST\_JUMP\_POS, VERY\_BRIGHT\_NEIGHBOR, SUSPECT\_RV\_COMBINATION.} and ANDFLAGS\footnote{Excluding stars with any of the flags: BAD\_PIXELS, BRIGHT\_NEIGHBOR, COMMISSIONING, PERSIST\_HIGH, PERSIST\_JUMP\_NEG, PERSIST\_JUMP\_POS, VERY\_BRIGHT\_NEIGHBOR.} to remove stars with problematic spectra. Moreover, we apply a cut in the ${\rm \chi^2_{ASPCAP}<20}$ (the $\chi^2$ quantifying the quality of the match of the observed spectrum with its best synthetic template) to remove stars whose $\chi^2$ places them at the poor-fitted-model tail of the $\chi^2$ distribution of the whole sample. We use ASPCAPFLAGS\footnote{Excluding stars flagged with STAR\_BAD.} to select stars not flagged as having an unreliable ASPCAP analysis. We also apply a cut in \logg~ selecting stars with \logg>0.5~dex, to exclude stars of low gravity and temperature whose spectra are more complicated to analyze. Finally, we apply a cut to keep stars with S/N>65. The ASPCAP pipeline provides results which are already of high quality at S/N=50 \citep{garcia-perez2016}, so we judge this cut as being conservative, removing a few potentially lower-quality spectra.  By using this combination of flags and cuts, we cull a robust sample comprising 86,057 stars.

\section{Distances and sample selection}
\label{sec:sample_selections}

We estimated spectro-photometric distances for the whole set of selected stars. To this end, we adopted their calibrated fundamental parameters \teff, \logg, metallicity \feh~ and the procedure described in \citet{rojas-arriagada2017}.

We can summarize this procedure as follows. The set of fundamental parameters and associated 2MASS $J$, $H$, $K_S$ photometry (and their errors) are adopted to simultaneously compute the most likely line-of-sight distance and reddening by isochrone fitting with a set of PARSEC\footnote{Available at \url{http://stev.oapd.inaf.it/cgi-bin/cmd}.} isochrones. To this end, we consider a set of isochrones\footnote{In practice, each isochrone consists of a sequence of model stars located along a track of increasing mass in the \teff \ vs. \logg~plane, going from the main-sequence to the AGB. Each model star is characterized by theoretical values of the absolute magnitudes $M_J$, $M_H$, and $M_K$.} spanning ages from 1 to 13 Gyr  and metallicities from -2.2 to 0.5 dex in steps of 1 Gyr and 0.1 dex, respectively. A given observed star can be placed in the theoretical/isochrone \teff-\logg-metallicity space, and so the parameter-space-distances $d_{iso}$ from this star to the whole set of model stars considering all the isochrones can be computed. From this set of parameter-space-distances we can compute weights to estimate predicted physical properties of the observed star. However, some considerations should be taken into account:
\begin{itemize}
\item We must account for the evolutionary speed of the model stars along the isochrone. The isochrones are constructed with their model stars  uniformly distributed along them. This means that a simple unweighted statistic using all the model stars will lead to overweight short evolutionary stages to the detriment of long-lived ones. We correct for this effect by including a weight $P_m$ proportional to the mass difference between contiguous model stars, in order to assign more weight to the long-lived evolutionary stages.
\item We must account for the fact that in a stellar population, mass distribution is not uniform but follows the IMF. We correct for this effect by including a weight $P_{IMF}$ proportional to the relative number of stars located in the mass interval between consecutive model stars along the isochrone.
\end{itemize}

From these considerations, and the parameter-space-distances computed above, we compute a set of weights as $W_j=P_mP_{IMF}\left[e^{-d_{iso}} \right]$, with $j$ running over the set of model stars from all isochrones. Using these weights, we can calculate the theoretical likely values of the absolute magnitudes $M_J$, $M_H$, and $M_K$ of  the observed star as the weighted mean or weighted median of the respective values of all model stars from the whole set of isochrones. Finally, from these computed theoretical absolute magnitudes we can compute the line-of-sight distance of the observed star from its distance modulus.

The availability of distances allowed us to perform some cuts to select stars likely confined to the inner Galactic region. In particular, we used the cylindrical Galactocentric distance ${\rm R_{GC}=\sqrt{X^2+Y^2}}$ to select a sample of likely bulge stars as those with ${\rm R_{GC}<3.5~kpc}$. The distance to the Galactic center is adopted as ${\rm R_\odot=8.2~kpc}$ \citep{bland-hawthorn2016}. This sample is composed of 2789 stars. This relatively generous distance cut intends to account for the typical 20-35\% uncertainties of spectro-photometric distances while keeping a large-enough sample. We checked that all the results presented hereafter do not change if we adopt a more stringent cut of, for example, 2.5 or 2~kpc. The spatial distribution of the flag-culled APOGEE sample as well as that of the likely bulge selected sample is displayed in the upper panel of Fig.~\ref{fig:distances}.

\begin{figure}[]
\centering
\includegraphics[width=8.7cm]{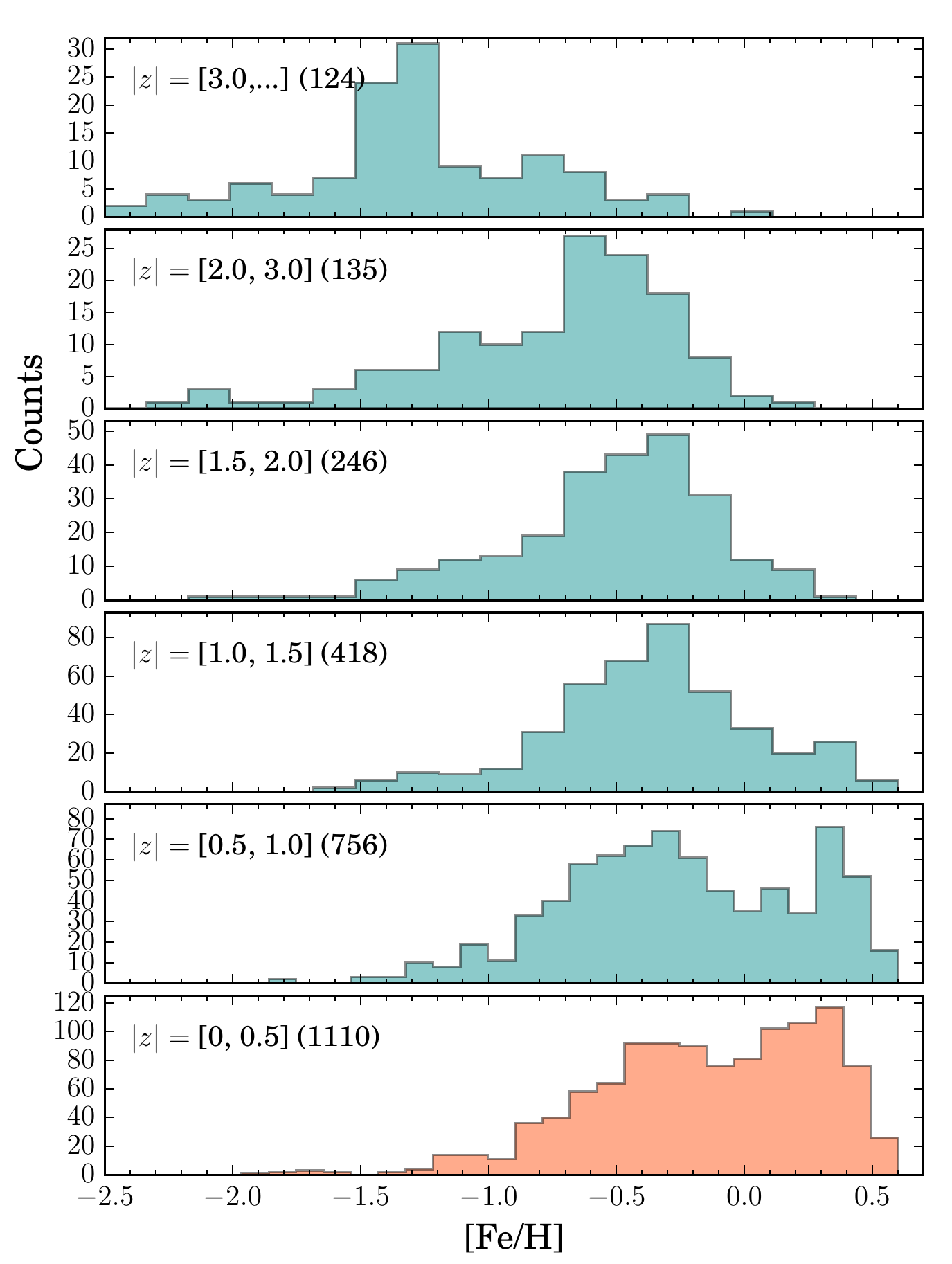}
\caption{Metallicity distribution function of ${\rm R_{GC}<3.5~kpc}$ stars in several bins of distance from the Galactic plane. The distance range of the considered bin and the number of stars falling in are quoted in each panel. The MDF corresponding to the close-to-the-plane sample (red box in Fig.~\ref{fig:distances}) is highlighted in red.} 
\label{fig:mdf_z}
\end{figure}

From the fact that APOGEE DR14 contains targets in a broad sky area (mostly at positive longitudes in the bulge region, lower panel of Fig.~\ref{fig:distances}), and considering that we have selected likely bulge stars based only on a cylindrical radial cut, we expect to find our stars spanning a broad range of distances from the Galactic plane. In fact, this is the case as seen in the upper panel of Fig.\ref{fig:distances}. Our bulge sample might contain a mix of bulge stars and a fraction of halo stars, increasing their relative importance far from the plane. To check for this, we divided the sample in a number of bins in absolute vertical distance from the plane. Their metallicity distribution functions are displayed in Fig.\ref{fig:mdf_z}. We can see a smooth transition from a bimodal MDF, dominated by metal-rich stars in the  bin closest to the plane, to a progressive dominance of metal-poor stars in the outer bins. Outside $|z|\sim3$~kpc, the remaining sample is dominated by halo stars as indicated by the single dominant peak at ${\rm [Fe/H]\sim-1.3~dex}$.

It is worth mentioning here that the relative proportion of metal-rich to metal-poor stars in the whole likely bulge sample might be biased. In fact, APOGEE targets observed towards the bulge are intrinsically bright stars, which are giant, cool, and red. Two effects might conspire to preferentially remove metal-rich stars from the final sample: First,  a bias might already exist imprinted in the input catalog. This might arise from the fact that the 2MASS input catalog becomes highly incomplete for the reddest stars in high-extinction regions due to the bright limit magnitude in the J-band. Metal-rich stars are intrinsically redder and are therefore more affected by the incompleteness in J. Second, intrinsically giant and cold stars have proven to be difficult to analyze with ASPCAP (as they are in general), especially when they are metal-rich. This might be the product of a mix of circumstances, the intrinsic difficulty in analyzing spectra with high spectral line crowding, continuum placement issues, and the fact that line opacity is not properly accounted for in the adopted atmospheric models when \teff~ goes below 3800~K \citep{meszaros2012}. In addition, there are not enough calibrators at high metallicity to produce a calibrated set of fundamental parameters and metallicity. For all these reasons, the APOGEE DR14 catalog does not contain calibrated parameters for cold giant stars. This cut might also preferentially remove metal-rich stars from the bulge sample analyzed here.

Although the MDF evolution with $|z|$ displayed in Fig.~\ref{fig:mdf_z} should be qualitatively correct, we refrain from making more quantitative assessments on the bulge MDF spatial variation. This will be thoroughly assessed in a detailed forthcoming work (Sobeck et al in prep). Instead, the following section focuses on the distribution of stars in the abundance--metallicity plane without taking density into account.

\begin{figure}[]
\centering
\includegraphics[width=9.2cm]{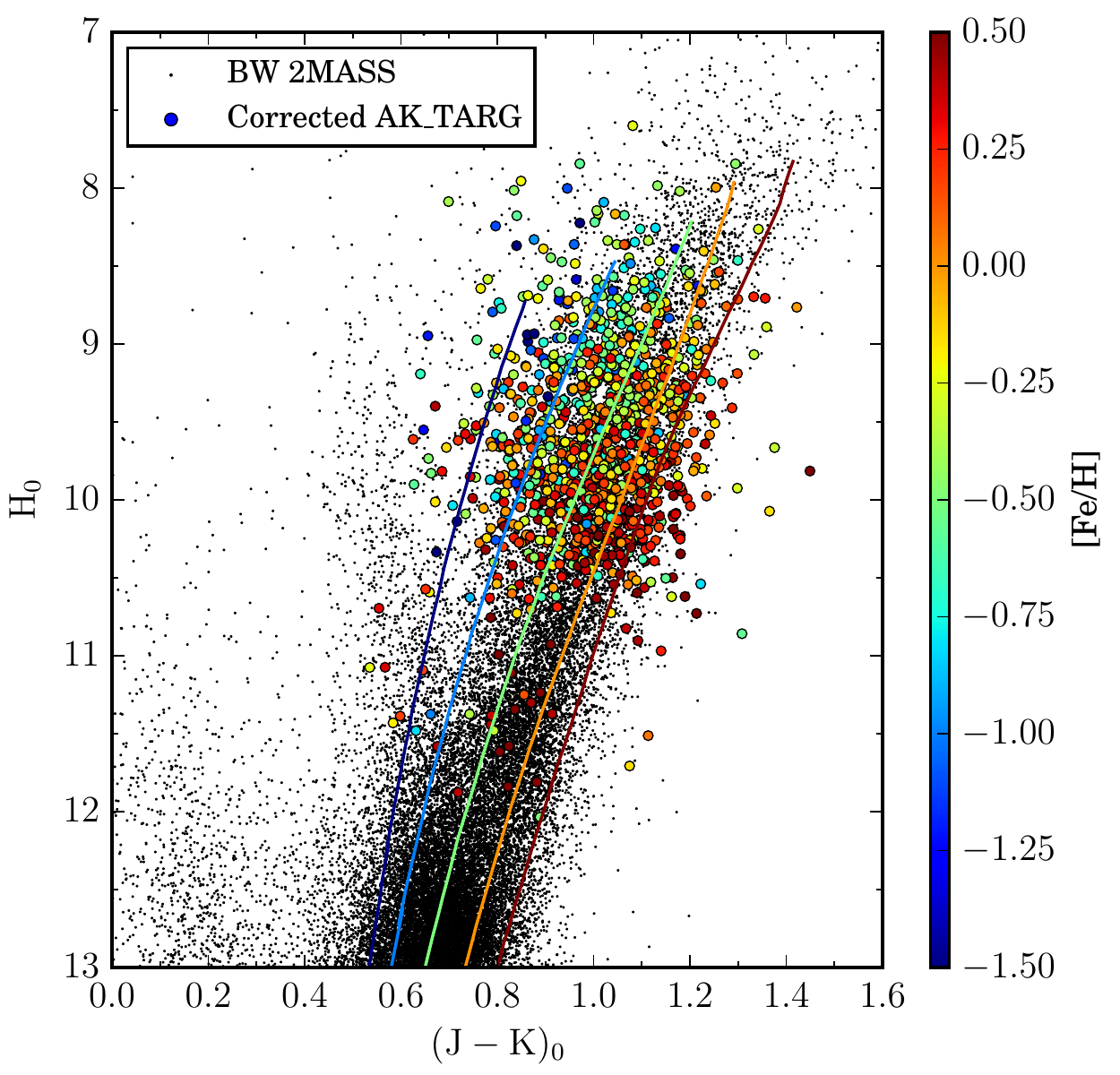}
\caption{2MASS $H_0$ vs $(J-K)_0$ diagram. Black dots represent stars in Baade's Window (as an example of a clean bulge-like CMD) dereddened using the extinction map \citet{gonzalez2012} through the beam calculator (\protect\url{http://mill.astro.puc.cl/BEAM/calculator.php}). Overlaid with circles and color coded according to their metallicity are the stars of the selected spectroscopic sample, dereddened by adopting the DR14 ancillary data for $A_K$. A set of PARSEC isochrones of 10 Gyr is depicted by the solid lines, also color coded according to their metallicity.}
\label{fig:cmd}
\end{figure}

\section{The double bulge sequence}
\label{sec:double_seq}

\begin{figure*}[]
\centering
\includegraphics[height=11cm]{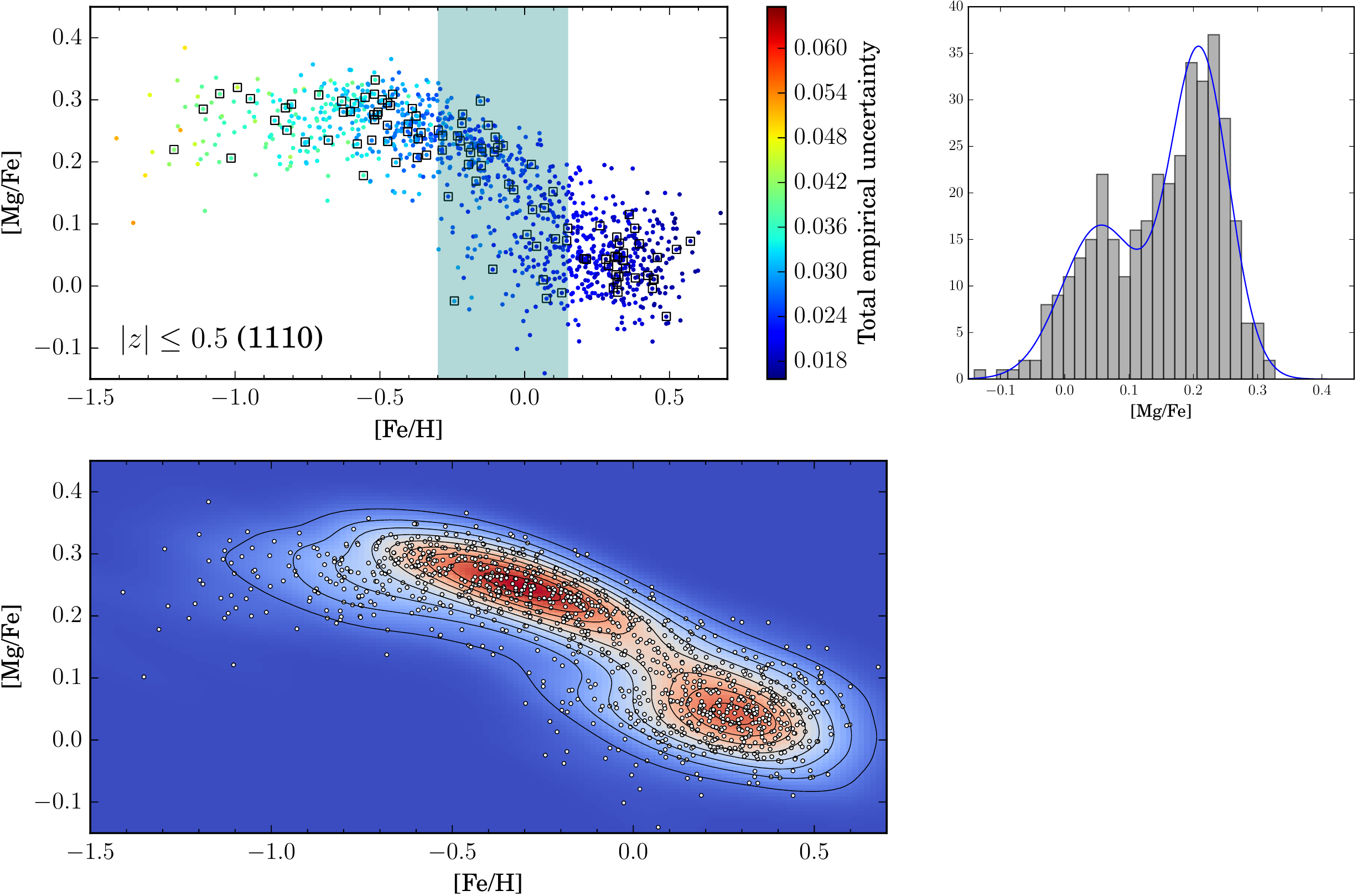}
\caption{\textit{Upper panels: }Distribution of close-to-the-plane bulge stars (${\rm R_{GC}}\leq3.5~kpc$, ${\rm |z|\leq0.5~kpc}$) in the [Mg/Fe] vs. [Fe/H] plane. Points are color coded according to their total empirical uncertainty defined as the sum in quadrature of the respective abundance measurement errors. Black squares highlight a subsample of stars for which relatively good-quality distance estimations (also indicating ${\rm R_{GC}}\leq3.5~kpc$) are available from the Gaia DR2 Bayesian distance catalog of \citet{bailer-jones2018}. A shaded vertical area highlights the metallicity range (${\rm -0.3\leq[Fe/H]\leq0.15}$ dex) in which stars are selected to plot their [Mg/Fe] ratio histogram in the right panel. The best (bimodal) Gaussian mixture, as estimated from an  analysis of Gaussian mixture models is displayed with a blue line. \textit{Lower panel: }Gaussian kernel estimation of the data (open black points) displayed as a density colormap, with a set of density contours as black lines.} 
\label{fig:seq_bimodal_Mg}
\end{figure*}

We want to examine the distribution of bulge stars in the abundance--metallicity plane. To this end we decided to focus on the closest-to-the-plane spatial bin (namely, constrained to ${\rm R_{GC}\leq3.5~kpc}$ and ${\rm |z|<0.5~kpc}$) whose MDF is displayed in the lower panel of Fig.~\ref{fig:mdf_z}, and whose $(l,b)$ distribution is shown in the lower panel of Fig.~\ref{fig:distances}. We choose this bin in order to obtain a similar proportion of metal-rich and metal-poor stars, and importantly to ensure that our sample is well confined in the bulge region. We are confident that this selection is not hiding possible spatial variations in the position of the sequence of bulge stars in the abundance--metallicity plane since such variations have not been detected so far in surveys examining specific pencil beam samples through the bulge area \citep[e.g.,][]{rojas-arriagada2017}.

In Fig.~\ref{fig:cmd} we display the 2MASS $H_0$ versus $(J-K)_0$ diagram of those close-to-the-plane sample stars; because they are distributed across a  patchy high-extinction region, we corrected them by their individual reddening vectors given by the $A_K$ values provided as ancillary data in the APOGEE DR14\footnote{From the AK\_TARG field, corresponding to the K-band extinction adopted for targeting.}. For comparison, we display with black points the 2MASS CMD of Baade's Window, as an example of a clean bulge-like CMD, dereddened by adopting the extinction map of \citet{gonzalez2012}. A set of PARSEC isochrones \citep{bressan2012} is overlaid, assuming an age of 10~Gyr and a distance to the Galactic center of 8.2~kpc \citep{bland-hawthorn2016}. These isochrones are color-coded by metallicity. We can see that, overall, our sample is well distributed along the red giant branch, spanning  the whole range of metallicity as depicted by the isochrones. Importantly, from this figure it is clear that the contamination by disk RC stars should be minimal. In fact, for the magnitude range of our sample, stars in the disk RC plume are well separated at bluer colors.

In the upper-left panel of Fig.~\ref{fig:seq_bimodal_Mg} we display with red points the distribution of our close-to-the-plane sample in the [Mg/Fe]-versus-[Fe/H] plane. The tight bulge sequence is relatively flat at low metallicity and seems to bend down at above ${\rm [Fe/H]=-0.4/-0.5~dex}$. Interestingly, there is a significant number of stars below the sequence from ${\rm [Fe/H]=-0.3~dex}$ up to ${\rm +0.1/+0.2~dex}$. The presence of these stars does not appear to be the product of a vertical shift of the bulge sequence above a given metallicity, but rather suggests the existence of a bimodality, with two parallel sequences at different [Mg/Fe] levels merging at supersolar metallicity. To further characterize this, in the right panel of Fig.~\ref{fig:seq_bimodal_Mg} we display the histogram of [Mg/Fe] for the stars in the metallicity range highlighted with the shaded green area in the left panel (${\rm -0.3\leq [Fe/H] \leq +0.15~dex}$, where the presence of the parallel sequences is visible). A clear bimodality in their density distribution confirms the presence of these two sequences with ${\rm [Mg/Fe]\sim0.05~and~0.25~dex}$. To further assess the statistical significance of the observed bimodality, we ran a Gaussian mixture models algorithm (GMM) over
this sample testing models from one to ten components. The data are well explained by a two-component solution as obtained from both the Bayesian and Akaike information criteria used for model selection. The selected model is overlaid with a blue line. Finally, in the lower panel of Fig.~\ref{fig:seq_bimodal_Mg} we display a Gaussian kernel estimation of the data to highlight the density structure. A set of contour lines is set for visual aid. From this panel, we can see how the low-Mg/metal-rich overdensity seems to be the locus where two sequences overlap: a sequence connecting this overdensity with that associated to the metal-poor MDF peak, and the sparse short sequence running at lower Mg abundance, as also seen in the contour line profiles.

It is worth emphasizing here that the presence of the parallel low-$\alpha$ sequence persists even if we adopt a more stringent cut in ${\rm R_{GC}}$ to define our sample of 2.5 or even 2~kpc (or if in an equivalent way, we limit the longitude range of our sample).

To perform this analysis, we selected magnesium because it is a very reliable element for performing chemical cartography. In fact, magnesium lines are strong enough to guarantee robust abundance determinations from spectra. They are, in addition, prone to only small nonlocal thermodynamic equilibrium (NLTE) effects/corrections, which can become important only at low metallicity \citep{asplund2005}, well below the metallicity range explored here. In addition, magnesium enrichment is largely dominated by core-collapse supernova (CCSN), in the sense that the yield from SN Ia is substantially small compared with that from massive stars \citep[][and references therein]{romano2010,rybizki2017}. Therefore, magnesium production can be thought of at first approximation as being due to a single nucleosynthetic channel.

In principle, the availability of several other chemical elements measured in APOGEE would offer us the opportunity to check the observed sequence structure, in particular using the whole set of $\alpha$-elements (especially Si and Ca). Unfortunately, the detailed examination of the behavior of measured abundances in APOGEE DR14 by \citet{zasowski2019} has revealed the presence of abundance trends with temperature for several elements. This latter study provides a detailed description of the trends and caveats concerning the use of elemental abundances produced by the ASPCAP pipeline for 11 elements. Of importance for our purposes, magnesium abundances are found to show no correlation with temperature across the entire metallicity range explored here. Similarly, \citet{jonsson2018} found that magnesium is the most accurate $\alpha$-element in APOGEE DR14 when comparing to optical analyses of a common set of stars.

\begin{figure}[]
\centering
\includegraphics[width=9.2cm]{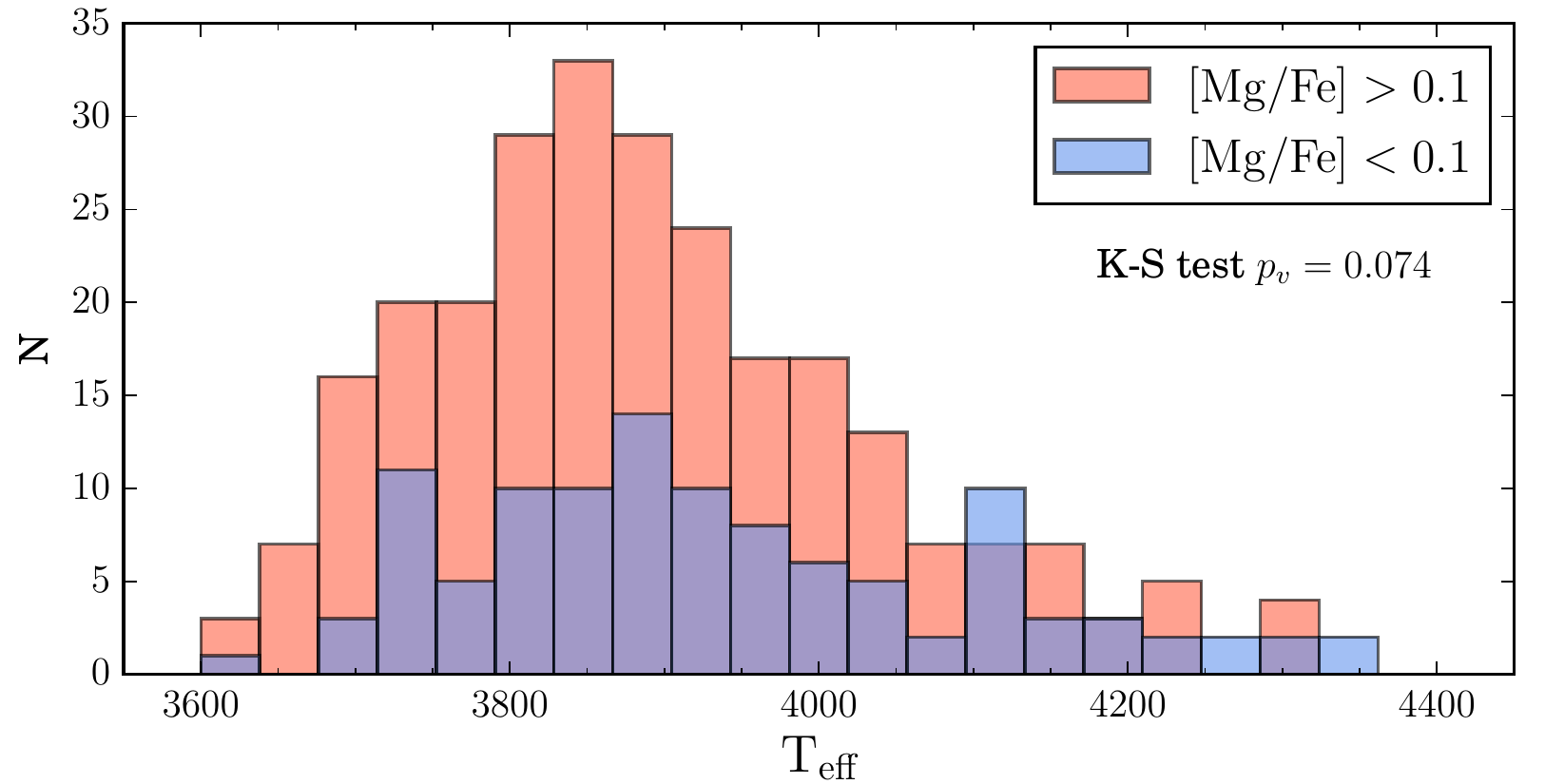}
\caption{\teff~ distribution for stars located in the metallicity region highlighted in Fig.~\ref{fig:seq_bimodal_Mg}, where the parallel sequences are observed. The red/blue histogram shows stars belonging to the high-/low-magnesium sequence. The p-value of a two-sample Kolmogorov-Smirnov applied on the date is quoted in the figure. From its value we cannot reject the null hypothesis that both distributions are drawn from the same parent population.}
\label{fig:teff_dist_bimodality}
\end{figure}

Even in the hypothetical situation in which a small, previously unnoticed abundance temperature correlation were found to exist, its effect concerning abundance differences between two samples would be nonetheless negligible as long as they had similar temperature distributions. To verify that, we display in Fig.~\ref{fig:teff_dist_bimodality} the temperature distribution of stars in the metallicity range where the Mg bimodality is detected (the shaded area in Fig.~\ref{fig:seq_bimodal_Mg}). The red/blue histograms correspond to high-Mg/low-Mg stars. From this figure, and the K-S test applied to the samples, we can see that stars in both sequences have temperature distributions which are comparable both in extent and relative density.

The general trends of the bulge sequence in the $\alpha$-abundance-versus-metallicity plane have already been characterized in the literature \citep{mcwilliam1994,fulbright2007,alves-brito2010,ryde2010,gonzalez2011,gonzalez2015,ness2013,rojas-arriagada2017,bensby2017}. Figure~\ref{fig:seq_bimodal_Mg} is novel in the sense that it presents for the first time a clear bimodality between the high-Mg sequence and the low-Mg group, which seems to include a metal-poor tail parallel to the high-Mg sequence. The magnesium abundance dispersion at supersolar metallicity seems to be larger than that of the metal-poor bulge sequence. This larger dispersion at supersolar metallicity could plausibly be explained by the overlap of the two different sequences, with the tight metal-poor one merging with the $\alpha$-poor one, starting from ${\rm [Fe/H]\rm-0.25~dex}$ towards supersolar metallicity.

\subsection{An independent distance assessment using Gaia DR2 data}
The Gaia mission \citep{perryman2001,gaia2016} promises to revolutionize our view of the Milky Way as it will provide a five- or six-parameter astrometric solution (positions, proper motions, parallax, and for a subset, radial velocity) for 1.3 billion stars, approximately 1\% of the stellar content of the Galaxy. The recently published second data release \citep[DR2,][]{gaia2018} is an important step in that direction, but currently, reliable distances are available for a limited (albeit unprecedentedly large) spatial volume. When considering stars further away than 4-5 kpc, for example, the still relatively large fractional errors in Gaia parallaxes (defined as PARALLAX\_ERROR/PARALLAX, using the respective quantities from the gaia\_source table\footnote{\url{https://gea.esac.esa.int/archive/documentation/GDR2/Gaia_archive/chap_datamodel/sec_dm_main_tables/ssec_dm_gaia_source.html}}) imply that reliable distances cannot be obtained by simply inverting the parallax. The biases that this simple estimation implies are thoroughly discussed in the Gaia literature \citep{luri2018}. A better approach is given by the adoption of Bayesian inference methods to get distances from parallaxes, which allow for reasonable distance estimations to be obtained even for stars with relatively large fractional parallax errors. A catalog of distances estimated through a Bayesian approach has been published for essentially the whole set of 1.33 billion stars contained in the Gaia DR2 catalog \citep[hereafter CBJ catalog,][]{bailer-jones2018}. These distances have been computed adopting as prior the exponentially decreasing space density prior presented in \citet{bailer-jones2015}. We refer the reader to \citet{bailer-jones2018} for technical details.

We used the CBJ catalog to perform an independent assessment concerning the spatial  distribution of the close-to-the-plane sample of stars, in particular those in the metallicity range where the bimodal sequence is seen (the shaded area in Fig.~\ref{fig:seq_bimodal_Mg}). As far-away sources, bulge stars are prone to presenting large fractional parallax errors. We cross-matched our close-to-the-plane sample with the CBJ catalog, keeping stars for which the fractional error is lower than 60\%. This ensures that Bayesian inferred distances are reasonable. We then selected stars which according to the CBJ distances also satisfy our definition of likely bulge stars (${\rm R_{GC}<3.5~kpc}$). In total, 110 stars meet these requirements; highlighted in the upper-left panel Fig.~\ref{fig:seq_bimodal_Mg} with black squares. In spite of their small number (10\% of the close-to-the-plane sample), we can see that they are evenly distributed with respect to the whole sample. Importantly, there are stars in the low-Mg sequence, showing an overall comparable picture, concerning the presence of the parallel sequences, to that found by using the whole close-to-the-plane sample as defined with spectro-photometric distances.

Although with a reduced number of stars, this assessment using Gaia distances is important since spectro-photometric and geometrical distances provide two independent measurements of the spatial distribution of our sample. In the case of the CBJ catalog distances, larger fractional parallax errors (even inside the conservative limits imposed to select our sample) might imply inferred distances likely more biased towards shorter distances than real ones. In this sense, the fact that stars in the low-Mg  sequence are inside the bulge region according to these geometrical distances represents a reinforcement of the general picture presented in Fig.\ref{fig:seq_bimodal_Mg}.

\subsection{Kinematics}
\label{subsec:kinematics}
In consideration of the two bulge sequences identified here, we split the whole close-to-the-plane bulge sample into two groups standing for the Mg-rich and Mg-poor sequences. To this end we adopted some simple cuts, defining the Mg-rich sequence as that given by the stars with ${\rm [Fe/H]\leq -0.3~dex}$ plus the high-Mg stars in the shaded area of Fig.~\ref{fig:seq_bimodal_Mg}, and the Mg-poor sequence as that given by those with ${\rm [Fe/H]\geq 0.15~dex}$ plus the low-Mg stars in the shaded area of Fig.~\ref{fig:seq_bimodal_Mg}.

\begin{figure*}[]
\centering
\includegraphics[width=18.4cm]{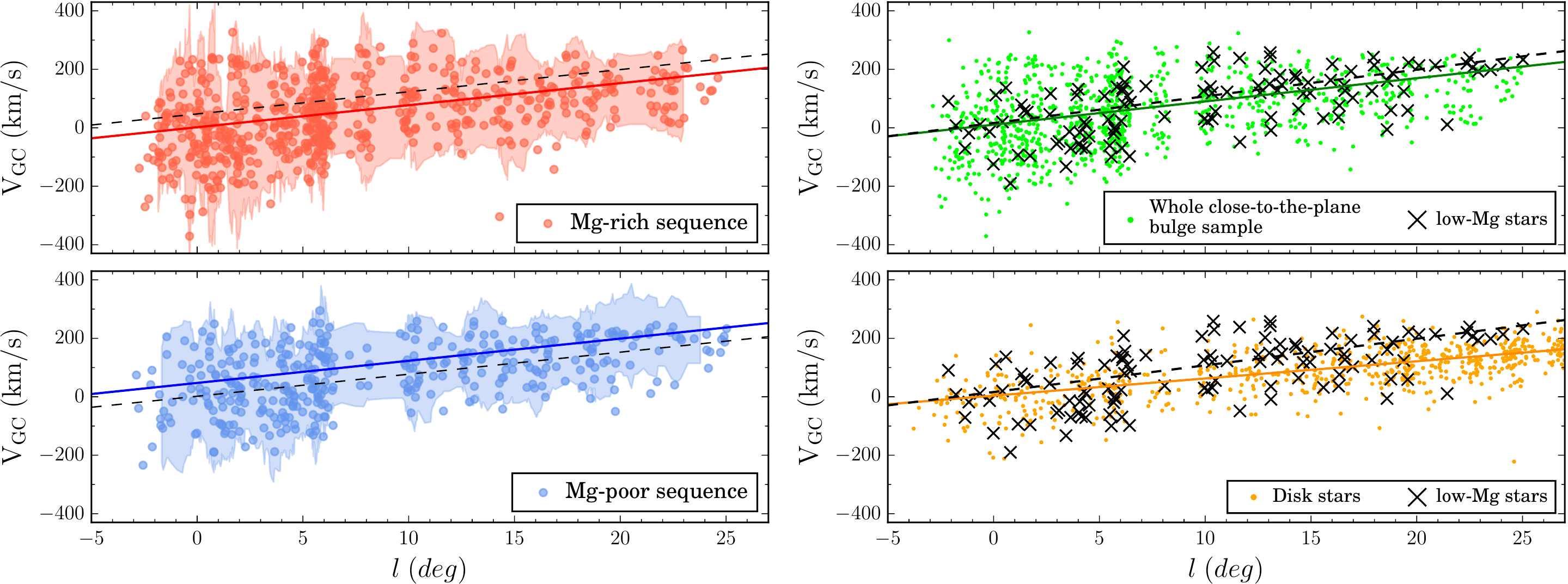}
\caption{Galactocentric velocity vs. longitude. \textit{Left panels:} Individual stars of the Mg- rich/poor sequences (as defined in the main text) are displayed with red/blue color in the upper/lower panel. In each case, the mean linear trend of the data is displayed as a solid colored line (estimated with the Theil-Sen estimator for robustness), and the $2\sigma$ dispersion band as a shaded area. For visual aid, in each panel the linear trend of the other distribution is displayed as a dashed black line. \textit{Right upper panel:} Whole close-to-the-plane sample (thus the union of all stars displayed in the left panels) is displayed as green points with its linear trend as a solid line of the same color. Black crosses depict the subsample of low-Mg stars (in the metallicity range where Mg-bimodality is seen; see main text), with its linear trend displayed as a dashed black line. \textit{Right lower panel:} Sample of disk stars, selected as $3.5<R_{GC}<5.0$ kpc and $|Z|<0.5$kpc, displayed as orange points, with its linear fit as a solid orange line. Black crosses and the dashed black line have the same meaning as in the upper panel.} 
\label{fig:l_vh_curves}
\end{figure*}

As our sample is relatively small and sparsely spread over a range of latitudes (mainly in $|b|\leq 3^\circ$, see Fig.~\ref{fig:distances}), we cannot attempt a detailed comparison of the spatial variations of the kinematics of stars belonging to the Mg-rich and Mg-poor sequences. Instead, in the left panels of Fig.~\ref{fig:l_vh_curves} we display the general trend of the Galactocentric velocity\footnote{Conceptually corresponds to the line-of-sight radial velocity that would be measured by a stationary observer at the position of the Sun; this is calculated as $V_{GC}=V_{HC}+220\sin(l)\cos(b)+16.5\left[\sin(b)\sin(25)+\cos(b)\cos(25)\cos(l-53)\right]$, which is the heliocentric radial velocity corrected by the motion of the Sun relative to the LSR and the motion of the LSR relative to the Galactic center.} with longitude. A linear fit was performed on each distribution using the Theil-Sen estimator, which has the virtue of being a central tendency robust linear estimator. A $2\sigma$ dispersion band was also computed for each distribution. By comparing both panels, it is clear that Mg-rich/metal-poor and Mg-poor/metal-rich stars present the same degree of rotation. This can be seen by comparing the solid lines, which show the fit to the data, with the dashed ones depicting the trend found in the other panel. Striking is the fact that at this level of statistical resolution, both metal-poor and metal-rich stars rotate at about the same rate. In fact, the similar gradient found here for the longitudinal trend of $V_{GC}$ for metal-rich and metal-poor stars is discrepant with the proper-motion rotation curves presented by \citet{clarkson2018}. In that work, different rotation curves (from longitudinal proper motion $\mu_l$ and relative photometric parallax) are found for metal-poor and metal-rich dwarf stars, with the latter showing higher rotation amplitude and a steeper gradient (with a significant slope ratio of metal-rich/metal-poor stars $=3.70\pm0.68$). On the other hand, studies focusing in RR Lyrae stars as tracers of the metal-poor end of the bulge MDF (below the range usually explored with RGB/RC stars), show that these stars present null or negligible net rotation (both using radial velocities; \citeauthor{kunder2016} \citeyear{kunder2016}, and proper motions; \citeauthor{contreras-ramos2018} \citeyear{contreras-ramos2018}).

A similar exercise is performed in the right panels of Fig.~\ref{fig:l_vh_curves}, this time comparing three different samples: First, the whole close-to-the-plane sample is displayed in the upper right panel. Second, the low-Mg stars in the metallicity range where the parallel sequences are seen (the shaded area in Fig.~\ref{fig:seq_bimodal_Mg}) are displayed in both panels. And third, a sample of disk stars was selected from the original flag-culled APOGEE sample (see Sect.~\ref{sec:data}) to be in the region defined by $3.5<R_{GC}<5.0$~kpc and $|Z|<0.5$~kpc (see upper panel of Fig.~\ref{fig:distances}), thus sampling a region right outside the bulge as defined by our Galactocentric distance cut. These stars are displayed as orange points in the lower panel, with its linear fit as an orange solid line. We can see in the upper right panel of Fig.~\ref{fig:l_vh_curves} that the trend displayed by the low-Mg stars is as pronounced as that of the whole close-to-the-plane sample. Instead, in the lower right panel, the trend displayed by the disk sample has a small slope and a relatively small dispersion around the trend. In comparison, the low-Mg sample runs well above the sequence of disk stars, especially at larger galactic longitudes. These different kinematic trends characterizing low-Mg and disk sample stars reinforce the idea (as already seen from spectro-photometric and geometrical distances) of the low-Mg stars being actually located in the bulge region. In fact, if these stars were located in the disk but were incorrectly labeled as bulge stars because of incorrect/biased distances, then we would expect their kinematic trend to be comparable to that of the disk comparison sample rather than to the whole close-to-the-plane sample, which is not the case.

\begin{figure}[]
\centering
\includegraphics[width=9.1cm]{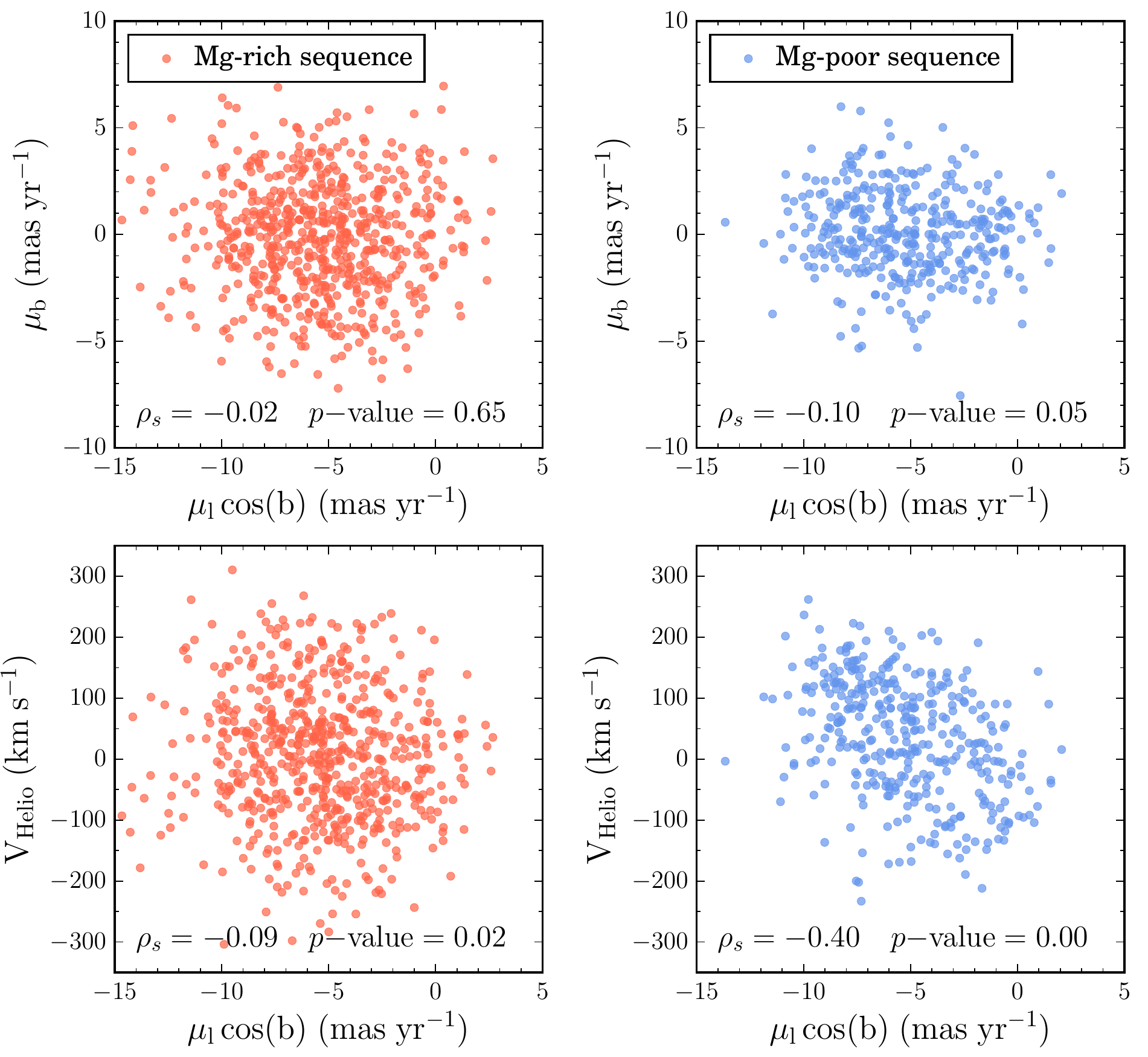}
\caption{\textit{Upper panels:} Vector point diagrams, from Gaia proper motions, of stars belonging to the Mg-rich and Mg-poor  sequences (left and right, respectively). \textit{Lower panels:} Longitudinal proper motion component vs. heliocentric radial velocity of stars belonging to the Mg-rich and Mg-poor sequences (left and right, respectively). In each panel, the Spearman correlation coefficient $\rho_s$ and its respective $p$-value are quoted. In all panels, mean error bars in proper motions and velocity are smaller than point sizes.} 
\label{fig:mul_mub_vhelio}
\end{figure}

From the cross-match of our catalog with the Gaia DR2, we obtained proper motions which are of good quality even for stars for which the parallax is not very informative. This dichotomy in the relative quality of proper motions versus parallax emerges naturally from the nature of the astrometric source model providing the five-parameter solution of Gaia sources \citep{luri2018}. In Fig.~\ref{fig:mul_mub_vhelio} we display the vector point diagrams (VPD) and the longitudinal proper motion versus heliocentric radial velocity,  for stars belonging to the Mg-rich and Mg-poor sequences separately. In each case, the spearman correlation coefficient and its $p$-value were computed to assess the significance of kinematic correlations. We can see that the VPD of the Mg-rich sequence is relatively round, with similar dispersion in both components of the proper motion. In contrast, Mg-poor sequence stars present a flatter distribution, with a smaller dispersion of latitudinal proper motion with respect to Mg-rich stars. In both cases, correlation coefficients are close to zero, indicating that the respective quantities are uncorrelated. A striking difference between the two groups of stars is seen in the lower panels of Fig.~\ref{fig:mul_mub_vhelio}, where Mg-rich stars show no correlation between heliocentric radial velocity and longitudinal proper motion (as indicated by $\rho_s\sim0$), while such a correlation exists for Mg-poor. In fact, the moderately large negative correlation coefficient plus a null $p$-value indicate a statistically significant negative correlation in this dataset.

The results of Figs.~\ref{fig:l_vh_curves} and ~\ref{fig:mul_mub_vhelio} argue for a different kinematic nature (and so, different orbital distributions) of stars belonging to the two bulge sequences reported here. Mg-rich stars display isotropic kinematics while Mg-poor star behavior is compatible with being a bar-supported population, kinematically colder in the vertical direction. The recognition of this kinematical dichotomy has already been characterized in the literature while separating samples based in metallicity only. Here we show that these results are fully recovered from our somehow more complex sample separation by recognizing the two bulge sequences in the [Mg/Fe]-versus-[Fe/H] plane.

\subsection{The [Mg/Fe] dispersion of the metal-poor bulge sequence}

\begin{figure}[]
\centering
\includegraphics[height=6.5cm]{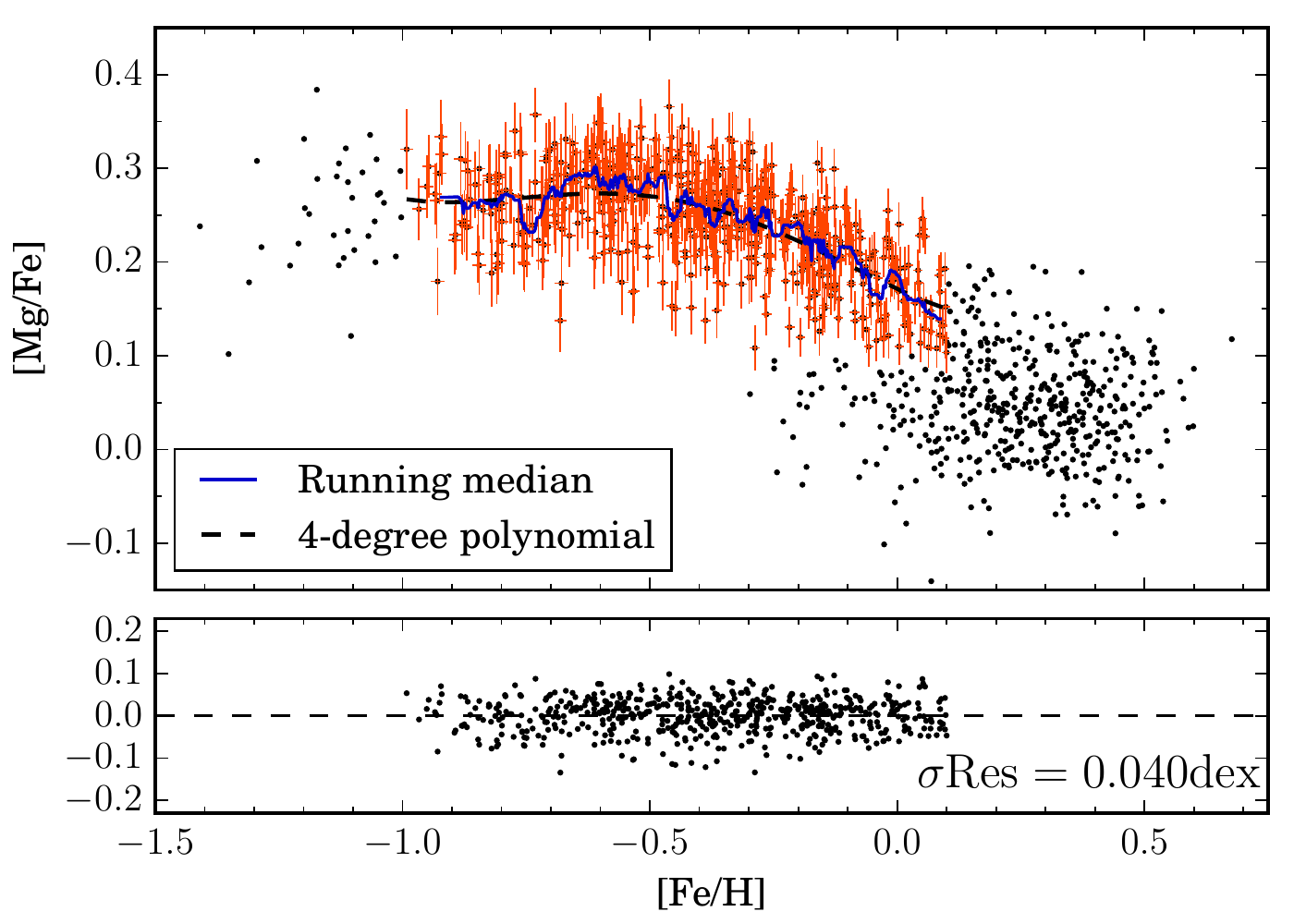}
\caption{Estimation of the [Mg/Fe] dispersion of the metal-poor bulge sequence in the [Mg/Fe] vs. [Fe/H] plane. \textit{Upper panel: } Black points stand for the whole close-to-the-plane bulge sample. The stars of the Mg-rich sequence with ${\rm[Fe/H]>-1.0~dex}$ are highlighted in red, with error bars accounting for their individual abundance measurement errors. A running median for these stars is displayed as a blue solid line. A fourth-degree polynomial model fit on the data is displayed by the black dashed line. \textit{Lower panel: }Residuals distribution from the polynomial fit. The dashed black line depicts the zero bias level. The standard deviation of the residuals is quoted in the panel.} 
\label{fig:a_disp}
\end{figure}

The specific distribution of stars of a given population in the $\alpha$-abundance-versus-metallicity plane provides important information concerning its formation history. In particular, the [$\alpha$/Fe] ratio dispersion of the stars around their mean sequence trend imposes limits to the extent of the IMF and chemical inhomogeneities involved in their formation. In fact, the plateau $\alpha$-level of a sequence depends on the IMF, and a small dispersion around this level is expected if the ISM from which the population was formed was sufficiently homogeneous, and/or the number of accreted stars is a small fraction.

In the upper panel of Fig.~\ref{fig:a_disp} we attempt a simple determination of the [Mg/Fe] dispersion (as a tracer of the general $\alpha$-abundance behavior) of the Mg-rich/metal-poor bulge sequence sample. In order to remove the influence of stars belonging to the metal-rich bulge, we keep stars with ${\rm [Fe/H]\leq0.1~dex}$ and ${\rm [Mg/Fe]\geq0.1~dex}$. We also ignore a small number of stars with ${\rm[Fe/H]<-1.0~dex}$. The selected sample of Mg-rich bulge sequence stars is highlighted in Fig.~\ref{fig:a_disp} as red points, with error bars accounting for their individual abundance measurement errors. We fit to these data a fourth-degree polynomial model, taking into account the individual [Mg/Fe] measurement errors. As can be seen, the fit follows the general distribution of the sample, which is also consistent with the nonparametric trend given by a running median (blue solid line) over the sample. In the lower panel of Fig.~\ref{fig:a_disp}, the residuals of our best fit to the data are displayed. We use them to estimate the mean [Mg/Fe] dispersion of bulge stars around their mean trend. The dispersion around the fit, as revealed by the distribution of the residuals, is relatively narrow and symmetric with a standard deviation of std=0.040~dex. The median of the individual [Mg/Fe] measurement errors is 0.026~dex. A subtraction in quadrature of both figures gives us an estimation of the astrophysical dispersion of bulge stars of ${\rm \sigma[Mg/Fe]_{bulge}=0.031~dex}$. To our knowledge, this is the first time that such an estimation is attempted. As discussed below, this small dispersion imposes conditions on the chemical homogeneity in the formation of the metal-poor bulge.

\section{Discussion and conclusions}
\label{sec:discussions_conclusions}

Here we present evidence supporting a bimodal double-sequence nature of the Galactic bulge in the [Mg/Fe]-versus-[Fe/H] plane. The last 10 years
of research has revealed the complex nature of the Galactic bulge. A clear bimodal (if not multimodal) nature of the bulge MDF has been claimed and further confirmed by a number of studies using data from some recent, large spectroscopic surveys, and also from  pencil beam samples in some specific low-extinction bulge windows \citep{hill2011,uttenthaler2012,ness2013,gonzalez2015,rojas-arriagada2017,schultheis2017,bensby2017,zoccali2017,garcia-perez2018}. The correlations with spatial distribution, kinematics, and ages for stars in the metal-poor and metal-rich modes of the MDF have revealed a distinct nature -- that is, distinct formation channels -- for them. Along these lines, there is general agreement in the literature concerning the origin of the metal-rich stars of the bulge, associating them with a population assembled by the secular evolution of the early disk via bar formation and buckling into an X-shaped structure. The origin of the metal-poor bulge remains debatable, with suggestions pointing to an early, fast episode of intense star formation \citep{hill2011,grieco2012,rojas-arriagada2017}, or the secular evolution of the thick disk \citep{ness2013,dimatteo2015,fragkoudi2018} to explain its formation.

Studying RC stars in Baade's Window, \citet{hill2011} found that the sequence in the [Mg/Fe]-versus-[Fe/H] plane might present a small vertical downshift in its trend at around solar metallicity. A more striking feature is suggested from the bulge data presented in \citet{recio-blanco2017}. In this latter study, a small subsample of bulge stars in Baade's Window from the Gaia-ESO survey\footnote{Unlike the bulk of bulge stars in the Gaia-ESO survey observed in one spectral window (HR21 GIRAFFE setup), this small subsample was observed using two of them (HR10 and HR21 GIRAFFE setups), which are those adopted for the main survey. This improved spectral coverage allowed for a more robust determination of fundamental parameters and elemental abundances. Consequently, some features, such as those discussed here, were not visible in the main bulge survey sample.} (GES) was examined to search for abundance-anomalous stars. In their Fig.~3 a small group of stars seem to draw a small parallel low-$\alpha$ sequence in the metallicity range ${\rm -0.4\leq [Fe/H]\leq -0.1~dex}$. Although their sample size is small, with only 48 stars, their $\alpha$-abundances are robust as they were determined from a large number of Mg, Ca, and Si clean optical lines. In this sense, the small tail of metal-rich low-$\alpha$ stars present in their sample, which is consistent with our results, seems to be robust. 

The red clump bulge sample of \citet[][from the Gaia-ESO survey]{rojas-arriagada2017} displays a sequence in the [Mg/Fe]-versus-[Fe/H] plane that appears to be relatively continuous over its whole metallicity range. In particular, the behavior of the chemical track at the high-metallicity end follows a general descending trend, with only a small hint of flattening to a constant [Mg/Fe] ratio. A similar pattern can be observed in other datasets, such as the bulge sample of \citet{gonzalez2015}, or in the larger sample (although based on lower-resolution spectra) of the ARGOS survey \citet{ness2013}. In contrast, the Mg-poor sequence of the close-to-the-plane sample is nearly flat, with a mean [Mg/Fe] enhancement which is close to but slightly larger than Solar value. The flattening of the bulge sequence in this metallicity range is also observed in the sample of 97 bulge microlensed dwarf stars of \citet{bensby2017}. The different behavior of the GES and ARGOS samples on one hand, and the \citet{bensby2017} and our close-to-the-plane sample on the other, could be due to differences in the details of the different adopted spectroscopic analysis strategies. From a chemical evolution point of view, the sequence drawn by GES data is consistent with a single track accounting for the chemical evolution of a population formed in situ, early and fast. As discussed in \citet{rojas-arriagada2017}, this scenario appropriately explains the shape of the sequence and the presence of the metal-poor mode of the observed MDF. A different complementary mechanism (secular evolution of the early thin disk)  must however be invoked to explain the presence of the metal-rich MDF peak. On the other hand, the trends seen in the \citet{bensby2017} and the close-to-the-plane samples provide a more explicit indication of a different mechanism/origin accounting for the metal-rich stars. In this context, the flattening of the trend can be taken as evidence of two different potential formation scenarios: This could be the result of a chemical evolution proceeding in a single track where alpha-element polluters become active again when  metallicity reaches near-solar levels. Alternatively, it could be an indication for the presence of two chemical tracks/sequences overlapping at high metallicity, one standing for the chemical evolution of the metal-poor stars formed in situ and promptly, and another accounting for stars formed over a longer timescale and rearranged into the bulge region through the secular evolution of the early thin disk. Although in the \citet{bensby2017} data there is no clear evidence of a parallel sequence, probably due to the small sample size, a general vertical jump at solar metallicity is evident, indicating different chemical regimes for metal-poor and metal-rich stars. On the other hand the age--metallicity relation of bulge stars found in that work \citep[as well as in][]{schultheis2017} is interesting in the context of the two sequences discussed here. In fact, the age distribution of metal-rich stars is wide, with a mixture of young and old stars. If, as we argue here, the two sequences of Mg-rich and Mg-poor bulge stars overlap at supersolar metallicity, then a mix of stars of different ages would be expected, with older stars coming from the Mg-rich sequence and younger ones belonging to the Mg-poor sequence.

From kinematics, we found that stars belonging to the two Mg-rich and Mg-poor bulge sequences seem to rotate at the same rate. This is in agreement with recent work examining the rotation curve of metal-rich and metal-poor bulge stars (as separated only from the MDF) across the bulge area \citep{ness2013,zoccali2017}. Nonetheless, this result shows a strong contrast with \citet{clarkson2018} where proper-motion rotation curves drawn by metal-rich and metal-poor dwarf stars are different in slope at a $5.4\sigma$ level. As pointed out in \citet{clarkson2018}, it seems unlikely that the different results from proper-motion- and radial-velocity-based rotation curves come uniquely from differences of giants and dwarfs as kinematical tracers. In fact, from the microlensed dwarf sample of \citet{bensby2017}, no significant differences in the mean radial velocity are seen between metal-rich and metal-poor stars. It seems more likely that differences come from the different way in which the bulge volume is sampled/projected while using both approaches. In particular, while the proper-motion rotation curves of \citet{clarkson2018} are based on a pencil beam sample, our rotation curves are drawn from a sample distributed over a wide area in $(l,b)$. A thorough investigation to elucidate this issue is needed but is out of the scope of this work. On the other hand, our results are in contrast with the observed kinematics of bulge RR Lyrae stars. Taking them as tracers of the old metal-poor component of the bulge, and studying both their radial velocities \citep{kunder2016}, and their proper
motions \citep{contreras-ramos2018}, it has been found that they show no net rotation. If they represent the low-metallicity end tail of the metal-poor distribution of old bulge stars, with their counterparts at higher metallicity given by the stars analyzed here, an explanation for their different kinematics is needed.

In summary, we present evidence arguing for the presence of a double sequence of bulge stars in the Mg-abundance-versus-metallicity plane. The spatial confinement of our sample to the bulge region is quantified using spectro-photometric distances, and is reinforced for a small subsample using geometrical distances obtained from  Bayesian inference of Gaia DR2 parallaxes with small fractional errors. An extra piece of information on the spatial location of the low-Mg stars in the bulge region is given from their similar kinematics with respect to the whole close-to-the-plane sample when comparing their run in the $V_{GC}$-versus-$l_{gal}$ plane. This shows that stars in the metal-rich and metal-poor modes of the bulge MDF represent in fact distinct evolutionary sequences that overlap at supersolar metallicity. In addition,  we estimated the astrophysical dispersion in ${\rm [Mg/Fe]}$ of the metal-poor sequence being as low as 0.033~dex.

In terms of formation scenarios of bulge components, if the Mg-rich sequence stars are the product of an in-situ formation, they might have been formed from a chemically homogeneous, well-mixed media, with the bulk of these stars formed at ${\rm [Fe/H]\sim-0.5/-0.4~dex}$ and a tail towards solar metallicity. In this context, the low [Mg/Fe] dispersion of 0.03~dex we found for this sequence considerably  limits the possibility that accretion played a relevant role in its formation, at least considering the chemistry of present day dwarf Milky Way satellite galaxies. As can be seen for example in Fig.~11 of \citet{tolstoy2009}, the $\alpha$-enhancement of dwarf galaxy stars at the metallicity range of metal-poor bulge stars would imply a large net [Mg/Fe] dispersion in the bulge sequence, which is not seen in our results. Concerning the Mg-poor sequence, the pieces of evidence presented and discussed here reinforce the case of these stars constituting an independent sequence which overlaps and outnumbers the Mg-rich one at supersolar metallicity. As a bulge population, these Mg-poor stars are the product of the secular evolution of the early thin disk.

Future data releases of APOGEE will provide better spatial coverage of the bulge region, as pointings at negative longitude are being observed as part of APOGEE-South. A larger number of stars will allow for more detailed studies of the properties of stars belonging to the two bulge sequences reported here. In particular, a higher spatial density of sources will enable us to study spatial variations of kinematics, and with the availability of age-CNO calibrations \citep[such as that of ][constructed for the APOGEE DR12]{martig2016}, will allow us to compare this with their age distribution. Needles to say that, as in many areas of astronomy, the future releases of Gaia, with astrometric solutions of increasingly high precision, will provide the opportunity to perform a joint analysis which might shed light on an unforeseeable number of open issues concerning the formation and evolution of the Galactic bulge.

\begin{acknowledgements}
ARA acknowledges partial support from FONDECYT through grant 3180203. ARB acknowledges financial support form the ANR 14-CE33-014-01. H.J. acknowledges support from the Crafoord Foundation, Stiftelsen Olle Engkvist Byggm\"astare, and Ruth och Nils-Erik Stenb\"acks stiftelse.\\

This work has been supported by the Ministry for the Economy, Development, and Tourism's Programa  Iniciativa  Cient\'\i  fica  Milenio through  grant IC120009,  awarded  to Millenium Institute of Astrophysics (MAS), the BASAL CATA Center for Astrophysics and Associated Technologies through grant AFB 170002,  and from FONDECYT Regular 1150345 and 1170121. \\

Funding for the Sloan Digital Sky Survey IV has been provided by the Alfred P. Sloan Foundation, the U.S. Department of Energy Office of Science, and the Participating Institutions. SDSS acknowledges support and resources from the Center for High-Performance Computing at the University of Utah. The SDSS web site is \url{www.sdss.org}.\\

SDSS is managed by the Astrophysical Research Consortium for the Participating Institutions of the SDSS Collaboration including the Brazilian Participation Group, the Carnegie Institution for Science, Carnegie Mellon University, the Chilean Participation Group, the French Participation Group, Harvard-Smithsonian Center for Astrophysics, Instituto de Astrofísica de Canarias, The Johns Hopkins University, Kavli Institute for the Physics and Mathematics of the Universe (IPMU) / University of Tokyo, Lawrence Berkeley National Laboratory, Leibniz Institut für Astrophysik Potsdam (AIP), Max-Planck-Institut für Astronomie (MPIA Heidelberg), Max-Planck-Institut für Astrophysik (MPA Garching), Max-Planck-Institut für Extraterrestrische Physik (MPE), National Astronomical Observatories of China, New Mexico State University, New York University, University of Notre Dame, Observatório Nacional / MCTI, The Ohio State University, Pennsylvania State University, Shanghai Astronomical Observatory, United Kingdom Participation Group, Universidad Nacional Autónoma de México, University of Arizona, University of Colorado Boulder, University of Oxford, University of Portsmouth, University of Utah, University of Virginia, University of Washington, University of Wisconsin, Vanderbilt University, and Yale University.\\

This work has made use of data from the European Space Agency (ESA) mission
{\it Gaia} (\url{https://www.cosmos.esa.int/gaia}), processed by the {\it Gaia}
Data Processing and Analysis Consortium (DPAC,
\url{https://www.cosmos.esa.int/web/gaia/dpac/consortium}). Funding for the DPAC
has been provided by national institutions, in particular the institutions
participating in the {\it Gaia} Multilateral Agreement.

\end{acknowledgements}

\bibliographystyle{aa}
\bibliography{biblio}

\end{document}